\documentclass[showpacs,preprintnumbers,amsmath,amssymb]{revtex4}

\usepackage[dvips]{graphicx}
\usepackage{psfrag}
\usepackage{amsmath}
\usepackage{epsfig}
\usepackage{amstext}
\begin{document}
\title{Non-equilibrium Differential Conductance through a Quantum Dot in  a  Magnetic Field}
\author{A.C.Hewson}
\author{ J. Bauer}
\affiliation{Department of Mathematics, Imperial College, London SW7 2AZ, UK}
           \author{A.Oguri}
 \affiliation{Department of Material Science, Osaka City University, Sumiyoshi-ku,
Osaka 558-8585 Japan
}

\date{\today}

\begin{abstract}
We derive an exact expression for the differential 
conductance for a quantum dot in an arbitrary magnetic field for small bias voltage. 
The derivation is based on the symmetric Anderson model using
renormalized perturbation theory and is valid for
all values of the on-site interaction $U$ including the Kondo regime. 
We calculate the critical magnetic field for the splitting of the Kondo resonance
to be seen in the differential conductivity as function of bias voltage.  
Our calculations for small field show that the peak position of the component 
resonances in
the  differential conductance 
are reduced substantially from  estimates using the equilibrium Green's
function. We conclude that it is important to take the voltage dependence of the local 
retarded Green's function into account in interpreting experimental results.
\end{abstract}
\maketitle

\section{Introduction} 
Measurements of the temperature dependence of the differential conductance
$dI/dV_{ds}$ through a quantum dot in the linear response regime have been shown to be in good in agreement
with  theoretical predictions  based on magnetic impurity models
 \cite{GGKSMM98,CHZ94}.
Measurements of $dI/dV_{ds}$ through a quantum dot in the presence of an applied 
magnetic field, however,  have revealed an apparent discrepancy between the
magnetic splitting in the Kondo peaks deduced experimentally and 
theoretical predictions based on the same models \cite{GSMAMK98,KAGGKS04,AGKK04}. 
The differential conductance measurements  are made in the presence of a
finite applied bias voltage $V_{ds}$, which is a situation for which there
are few theoretical predictions.  There are calculations based on the
non-crossing approximation (NCA) \cite{MWL93}. This method, however, is known to
break down in the Fermi liquid regime in the equilibrium case, so there is
some uncertainty about the validity of the predictions in this case. There
are also some recent perturbation calculations of Fuji and Uedo \cite{FU05} using
the Keldysh formalism, taking diagrams up to fourth order in the on-site
interaction $U$ into account. This approach should  reliably describe the
trends in the approach to the strong correlation limit, as the interaction
strength is increased, but cannot fully describe the Kondo regime, where
there is an exponential renormalization of the Kondo resonance width as a
function of $U$. Rosch et al. have calculated the differential conductance
using a perturbative renormalization group calculation for the exchange coupling
$J$ to leading order in $1/\mathrm{ln}(V_{ds}/T_{\rm K})$, which is
appropriate  for the very high voltage regime ($V_{ds}\gg T_{\rm K}$)
\cite{RPKW03} (see also \cite{GP05}).
This approach is, however, not suitable for investigating the Kondo regime,
where $V_{ds}$ and the magnetic field are of order of the Kondo temperature $T_{\rm K}$.

More accurate methods \cite{Cos00,Hof00,MW00,LD01} for the Kondo
regime exist for calculations in an applied magnetic field, which have been
used to interpret the experimental results \cite{GSMAMK98,KAGGKS04,AGKK04}, but they are
restricted to the  linear response (equilibrium) regime.
It is not clear that using the equilibrium response for $dI/dV_{ds}$ in the
presence of an applied bias voltage will lead to reliable predictions. 
In this paper we examine the effect of a finite applied bias voltage in the
small bias regime. In the limit of a small magnetic field $B$, we 
calculate asymptotically exactly the shifts of the peaks of the differential
conductance of particular spin component for small, but finite $V_{ds}$.
The results are based on renormalised perturbation theory and 
numerical renormalization group (NRG) calculations.
 We also deduce
the value of the critical field $B_c$ for two distinct peaks to be seen in
$dI/dV_{ds}$. The knowledge of $dI/dV_{ds}$ to order
$V_{ds}^2$ for arbitrary values of $B$ is sufficient to determine $B_c$ exactly.

In our treatment the quantum dot will be described by a
single impurity Anderson model with particle-hole symmetry coupled  via left
and right leads to two reservoirs of free electrons. The  Anderson model  has
the form, 
\begin{equation} 
H_{\rm AM}=\sum\sb {\sigma}\epsilon\sb {d\sigma}
d\sp {\dagger}\sb {\sigma}
d\sp {}\sb {\sigma}+
Un\sb {d,\uparrow}n\sb {d,\downarrow}
 +\sum\sb {{ k},\sigma}( V\sb { k}d\sp {\dagger}\sb {\sigma}
c\sp {}\sb {{ k},\sigma}+ V\sb { k}\sp *c\sp {\dagger}\sb {{
k},\sigma}d\sp {}\sb {\sigma})+\sum\sb {{
k},\sigma}\epsilon\sb {{ k}}c\sp {\dagger}\sb {{ k},\sigma}
c\sp {}\sb {{
k},\sigma},
\label{ham}
\end{equation}
where $\epsilon_{d\sigma}=\epsilon_d -\sigma g\mu_{\rm B}B/2$ is the energy of
the impurity levels with spin $\sigma=\pm 1$ 
in a magnetic field $B$, $U$ is the interaction at the impurity site,
 and $V_{k}$ the hybridization matrix element to a band of conduction electrons with
energy $\epsilon_k$. In the wide band limit the hybridization weighted density of states,
 $\Delta(\omega)=\pi\sum_{k}|V_k|^2\delta(\omega-\epsilon_k)$, can be taken as
 a constant $\Delta$. Taking this limit justifies the neglect of any magnetic
 field term acting on the conduction electrons in (\ref{ham})
 (Clogston-Anderson compensation theorem \cite{CA61}).
Any polarization of the conduction electrons only affects the impurity via
the hybridization function $\Delta(\omega)$ and
any change to the conduction band density of states due to an applied magnetic
field is only at the band edges ($\pm D$), and hence negligible if $g\mu_{\rm B} B\ll D$.
Therefore, if $\Delta(\omega)$ is independent of $\omega$ the magnetic field does
not need to be included in the electron band.

The single impurity model can be applied to a quantum
 dot which is equally coupled via leads to left and right reservoirs. In this
 situation the charge on the quantum dot is coupled only to an even
 combination of states from the left and right channels by an effective
 hybridization $V_k$, so it can be mapped into a single channel model. If
 there is a potential difference $eV_{ds}$ due to a bias voltage $V_{ds}$ ($e$
 is the electronic charge), then the chemical potentials in the left and right 
reservoirs differ from the average chemical potential $\mu$ by $eV_{ds}/2$ and
$-eV_{ds}/2$, respectively.\par 
A general expression for the current $I$ due to the applied voltage 
through a quantum dot has been given by Hershfield et al. \cite{HDW91} and Meir
and Wingreen  \cite{MW92,WM94}, and specialising it to this 
symmetric case it takes the form, 
\begin{equation}
I=\frac{e\Delta}{h}\sum_\sigma\int\limits_{-\infty}^{\infty}{d\omega}
[f_L(\omega)-f_R(\omega)][-{\rm Im}G^r_{d\sigma}(\omega, V_{ds})],
\label{diffcond}
\end{equation}
where $G_{d}^r(\omega,V_{ds})$ is the steady state retarded Green's function
on the dot site, and $f_L(\omega)$, $f_R(\omega)$ are Fermi distribution
functions for the electrons in the left and right reservoirs, respectively,
and $h$ is Planck's constant.\par 
The retarded Green's function 
$G_{d\sigma}^r(\omega, V_{ds})$ can be written in terms of a self-energy $\Sigma_\sigma(\omega,V_{ds})$,
 \begin{equation}
G_{d\sigma}^r(\omega,V_{ds})=\frac{1}{\omega-\epsilon_d+i\Delta-\Sigma_{\sigma}(\omega,V_{ds})}
\end{equation}
Oguri \cite{Ogu05,Ogu01} has shown that this Green's function for 
the symmetric model ($\epsilon_d=-U/2$) in the Fermi liquid regime in the absence of a magnetic
field to order $\omega^2$ and $(eV_{ds})^2$,  
can be expressed in the form,
\begin{equation}
G_{d\sigma}^r(\omega,V_{ds})=\frac{z}{\omega+i\tilde\Delta-\tilde\Sigma(\omega,V_{ds})}
\end{equation}
where $\tilde\Sigma(\omega,V_{ds})$ is a renormalized self-energy \cite{Hew93} given by 
\begin{equation}
\tilde\Sigma(\omega,V_{ds})=-i
%{2\tilde\Delta}\left(\frac{\tilde  U}{\pi\tilde\Delta}\right)^2
c\left[\omega^2+{3\over  4}(eV_{ds})^2\right],
\qquad {\rm with}\;\;
c=\frac1{2\tilde\Delta}\left(\frac{\tilde  U}{\pi\tilde\Delta}\right)^2 
\label{self1}
\end{equation}
expressed in terms of the renormalized parameters $\tilde\Delta$ and $\tilde
U$. These parameters \cite{Hew93} are defined in terms of the local impurity
self-energy $\Sigma(\omega)$ and local irreducible 4-vertex
$\Gamma_{\uparrow\downarrow}(\omega_1,\omega_2,\omega_3,\omega_4)$
\cite{Hew93} for $V_{ds}=0$, 
\begin{equation}
\quad\tilde\Delta =z\Delta ,\quad \tilde U=z^2\Gamma_{\uparrow\downarrow}(0,0,0,0),
\label{ren}
\end{equation}
where $z$ is given by
$z={1/{(1-\Sigma'(0))}}$. The renormalized parameters $\tilde\Delta$ and $\tilde U$,
can be calculated directly from NRG calculations,
as described in references \cite{HOM04,Hew05}, where they are given in terms of the 'bare'
parameters of the model $\Delta$ and $U$. Alternatively they can also be deduced
  from the Bethe ansatz results \cite{Hew93}.\par  

\begin{figure}
\begin{center}
\includegraphics[width=0.6\textwidth]{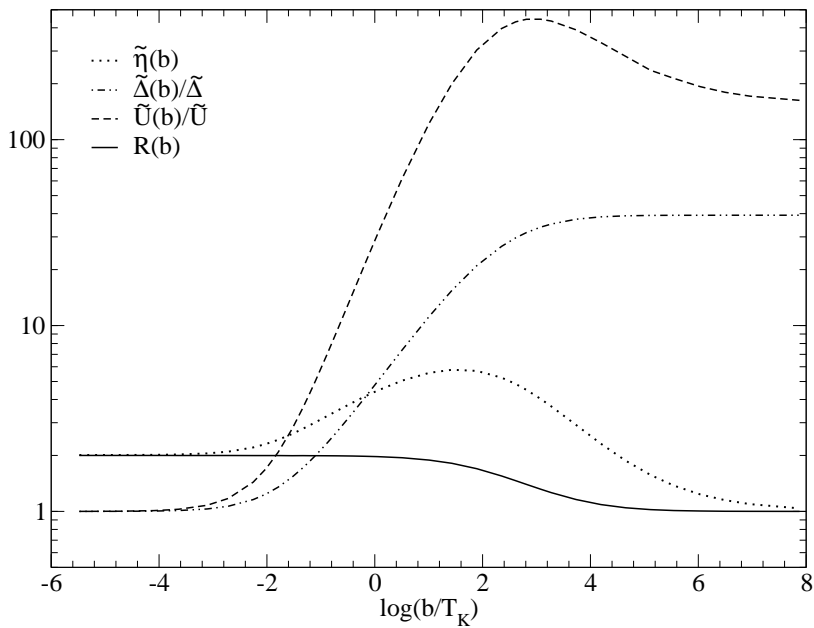}
\smallskip
\end{center}
\caption{$b$ dependence of the renormalized parameters for strong coupling
  ($U/\pi\Delta=4$), in relation with their
  $b=0$ values, $\tilde U=\pi\tilde\Delta=4T_{\rm K}=0.0817\Delta$. Note the
  logarithmic scale of the y-axis.}
\label{renpar}
\end{figure}

We can generalise this result by including a magnetic field term
\cite{Hew05,HBK05} to lowest order in $B$, and then $G_{d\sigma}^r(\omega)$
takes the form,  
\begin{equation}
G_{d\sigma}^r(\omega)={z\over\omega+\sigma\tilde\eta
    b+i\tilde\Delta-\tilde\Sigma_{\sigma}(\omega,V_{ds})},
\label{gf1}
\end{equation} 
where $\sigma=\pm 1$, $b=g\mu_{\rm B}B/2$, and $\tilde\eta$ is a
renormalization parameter defined as
$\tilde\eta(b)=z(b)(1-\Sigma_{\uparrow}(0,b)/b)$.
This renormalized parameter  can  be deduced from NRG calculations as shown in
reference \cite{Hew05,comment}. We give an example in the Kondo regime 
for $U/\pi\Delta=4$ in figure \ref{renpar}, where the ratio of the parameters 
to their values in zero field are plotted as a function of ${\rm log}(b/T_{\rm
  K})$. As $b\to 0$,
$\tilde\eta(b)$ is independent of $b$ such that $\tilde\eta(b)\to R$, 
where $R$ is the Wilson or '$\chi/\gamma$' ratio for zero magnetic field.  In
terms of the renormalized parameters for finite field $R(b)$ is given by $R(b)=1+\tilde
U(b)\tilde\rho^{\scriptscriptstyle 0}_d(0,b)$ \cite{Hew93,HOM04}, where the free quasiparticle density of
states $\tilde \rho^{\scriptscriptstyle 0}_{d,\sigma}(\omega,b)$ is given by
\begin{equation}
\tilde\rho^{\scriptscriptstyle 0}_{d,\sigma}(\omega,b)={\tilde\Delta(b)/\pi\over
    (\omega+\sigma b\tilde\eta(b))^2+\tilde\Delta^2(b)}, 
 \end{equation}
which is independent of the spin index $\sigma$ for $\omega=0$. 
  There are in general  $B^2$ and higher order terms in the self-energy
but for the moment  we  work only to first order in $B$ so the
renormalized self-energy $\tilde\Sigma_\sigma(\omega,V_{ds})$ 
is still given by equation (\ref{self1}) with 
\begin{equation}
c=c(b)=\frac{\pi\tilde
  U^2(b)[\tilde\rho^{\scriptscriptstyle 0}_{d}(0,b)]^3}{2},   
\label{cb}
\end{equation}
which reduces to (\ref{self1}) for $b\to 0$.
\section{Weak field limit} 
\begin{figure}[h]
\unitlength=1cm
      \begin{picture}(16,4)
        \put(1,0.5){
    {\resizebox{6cm}{3cm}{\epsfig{file=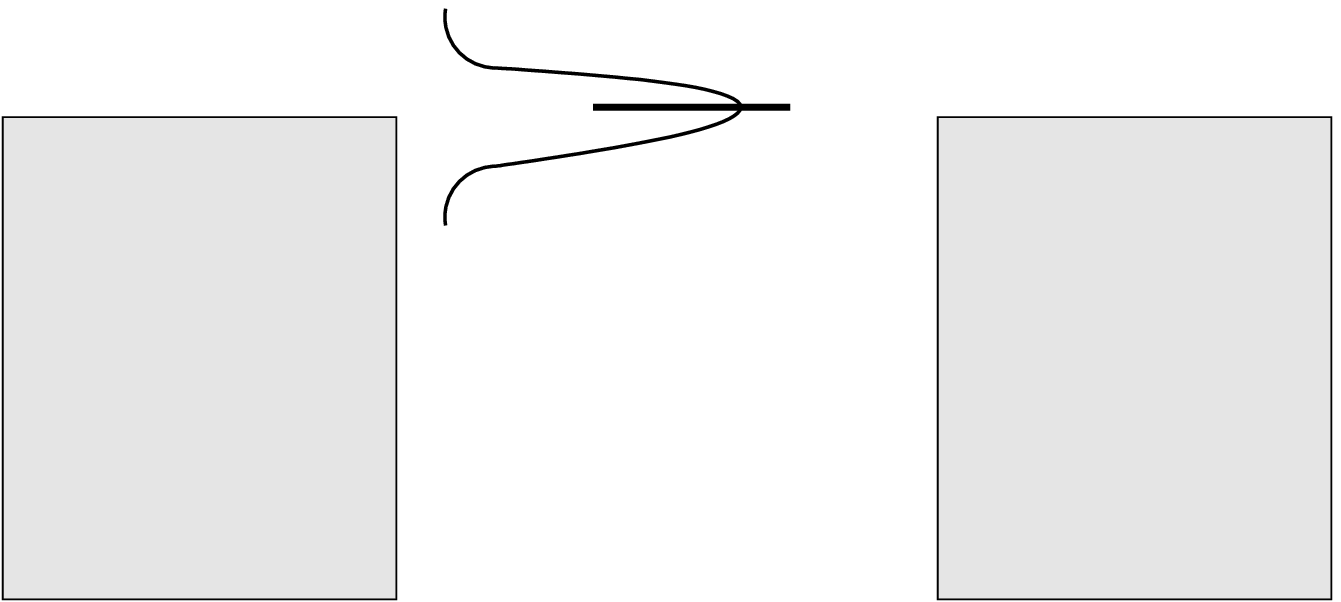}}}}
%\put(3.8,0.8){$U$}
\put(3.3,0.0){(a)}
\put(8,0.5){
    {\resizebox{6cm}{3cm}{\epsfig{file=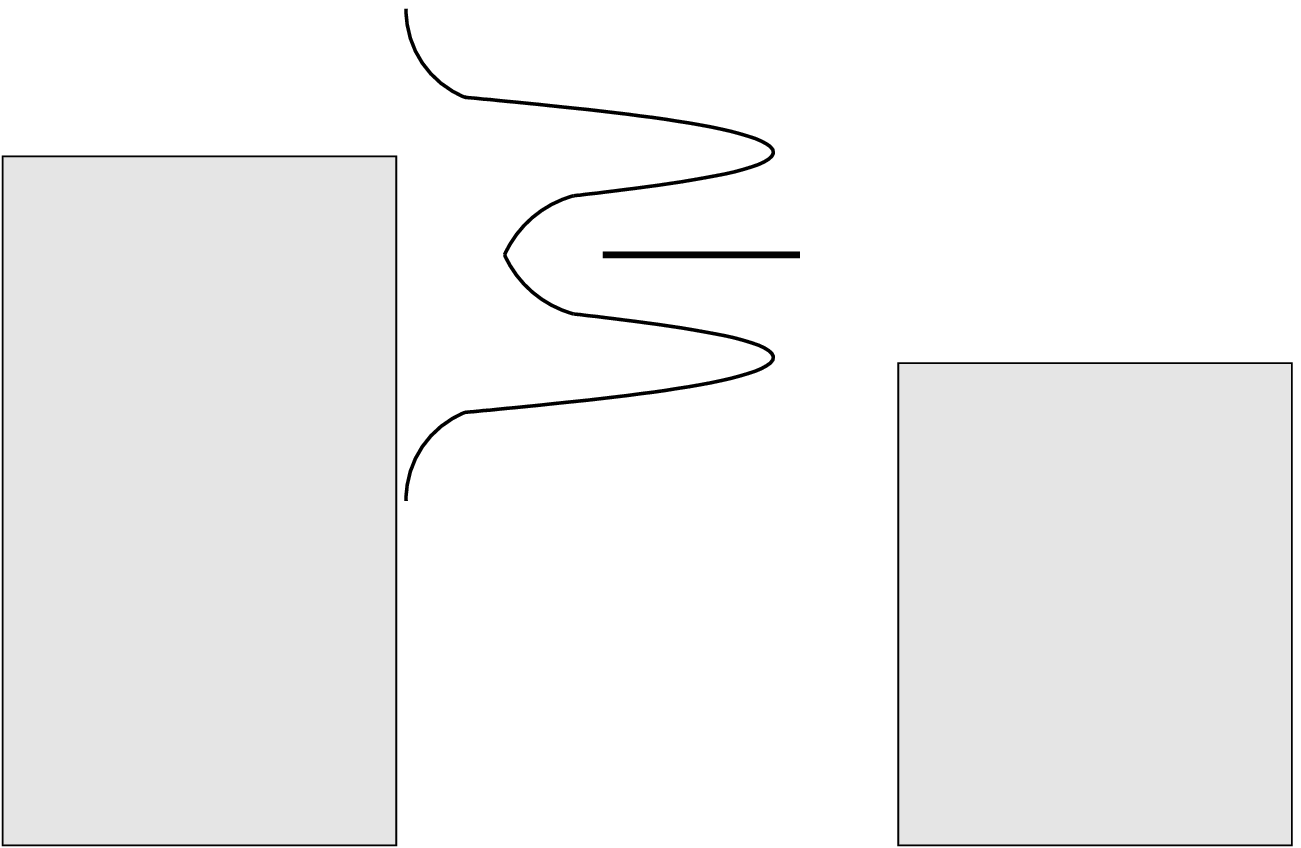}}}}   
\put(8,3.2){$\mu+eV_{ds}/2$}
\put(12.3,2.4){$\mu-eV_{ds}/2$}
\put(11.9,2.6){$\mu$}
%\put(11.7,0.9){$U$}
%\psfrag{mu+}{$\mu+eV_{ds}/2$}
%\psfrag{mu}{$\mu$}
%\psfrag{mu-}{$\mu-eV_{ds}/2$}
\put(10.5,0.0){(b)}
   \end{picture}
    \caption{A schematic  plot of the spectrum on the dot and chemical
      potentials for left/right lead ($\mu\pm eV/2$) and dot ($\mu$) for (a)
      zero bias and zero magnetic field and (b) finite voltage and finite field.}
\label{chempot}
\end{figure}

We look at the effects of the different contributions to $dI/dV_{ds}$ in turn.
We consider the simplest situation to begin with and ignore the renormalized
 self-energy
term in (\ref{gf1}). The differential conductance (\ref{diffcond}) at $T=0$
then takes the form,
\begin{equation}
{dI\over dV_{ds}}={e^2\over h}\sum_\nu{\tilde\Delta^2\over
    (eV_{ds}/2-\nu\tilde\eta b)^2+\tilde\Delta^2},
\label{didv1}
\end{equation}
where $\nu=\pm 1$. The total result
is expressed as a sum over $\nu=\pm1$, rather than  as a sum over
$\sigma=\pm 1$, as both spin up and spin down states
 contribute to each resonance. Note the result does not simply correspond to the
non-interacting case, as it includes the many-body renormalization factor
$\tilde\eta$. It corresponds to the case in which the interaction between the
renormalized quasiparticles ($\tilde U$) is neglected. 
In the Kondo regime $\tilde\eta=2$,
so the effective Zeeman splitting of the resonance in this limit is twice the
Zeeman splitting ($2b$) for non-interacting electrons ($U=0$).  
Note that the energy level on the dot is at the average chemical
potential $\mu$, which is shifted by  $eV_{ds}/2$ with respect to the value at
zero bias. 
The individual terms in (\ref{didv1}) for each $\nu$ have maxima at
$\pm 2\tilde\eta b$. The maxima of the sum in general occur at a
shifted position $eV_{ds}^{\pm}=\pm 2\tilde\eta b f_c(\tilde\eta b,
\tilde\Delta)$ with 
\begin{equation}
f_c(\tilde\eta b,\tilde\Delta)=
\left[1-\left(1-\left[1+\left(\tilde\Delta/\tilde\eta
        b\right)^2\right]^{1/2}\right)^2\right]^{1/2},
\label{fc}
\end{equation} 
provided that $\tilde\eta b > \tilde\Delta/\sqrt{3}$. For $\tilde\eta b \gg
\tilde\Delta$ can be approximated by $f_c(\tilde\eta 
b,\tilde\Delta)\simeq 1-\frac18\frac{\tilde\Delta^4}{(\tilde\eta  b)^4}$. 
The peak splitting is therefore $e(V_{ds}^{+}-V_{ds}^{-})=4\tilde\eta b
f_c(\tilde\eta b,\tilde\Delta)$.    
A maximum of the differential conductance, therefore, occurs when one of the 
quasiparticle peaks  is coincident with 
left Fermi level at $\mu+eV_{ds}/2$ and at the same time the other peak
coincides with the right Fermi level, $\mu-eV_{ds}/2$.  This is illustrated in figure \ref{chempot}.
\noindent 
In the interpretation of the experimental results of $dI/dV_{ds}$ the
splitting of the Kondo resonance $\Delta_{\rm Kondo}$ was identified with the voltage 
splitting $e(V_{ds}^+ -V_{ds}^-)=\Delta_{\rm Kondo} $\cite{KAGGKS04,AGKK04}. However, for the
renormalized quasiparticles the Kondo splitting is $\Delta_{\rm Kondo}=2\tilde
\eta b$, which substituted in the above result gives $e(V_{ds}^+
-V_{ds}^-)=2\Delta_{\rm Kondo}$. The factor of 2 difference is due to the
many-body shift in the energy level on the dot, which was not taken into
account in the interpretation of the experimental results in \cite{KAGGKS04,AGKK04}.

We should also take into account the terms in the renormalized self-energy
 $\tilde\Sigma(\omega,V_{ds})$. For the moment we ignore the $V_{ds}$
 dependence of $\tilde\Sigma(\omega,V_{ds})$, and then 
 \begin{equation}
{dI\over dV_{ds}}={e^2\tilde\Delta\over h}
   \sum_\nu\left.{\tilde\Delta-\tilde\Sigma^I(\omega)\over
     (\omega-\nu\tilde\eta
     b)^2+(\tilde\Delta-\tilde\Sigma^I(\omega))^2}\right|_{\omega=eV_{ds}/2},
\label{ft}
\end{equation}
where  $\tilde\Sigma^I(\omega,V_{ds})$ is the imaginary part of
$\tilde\Sigma(\omega,V_{ds})$ as given in equation (\ref{self1}). 
If the component spectral densities ($\nu\pm 1$) have maxima at $\omega_{\pm}$ then this expression
has maxima as a function of $V_{ds}$ for $2\omega_{\pm}$. Working to lowest order in $b$, we find from
the exact result for the self-energy to order $\omega^2$, that for a maximum, 
\begin{equation}
\omega=\pm\tilde\eta b+{\tilde\Delta\over
    2}{\partial\tilde\Sigma^I\over\partial\omega}=\pm\tilde\eta b-
 \omega\tilde\Delta c(b) 
% {\omega\over    2}\left(\tilde U\over \pi\tilde\Delta\right)^2.
\end{equation} 
The peak position in the resonances in the spectral
density to first order in $b$ is given by
\begin{equation}
\omega_{\pm}=\frac{\pm\tilde\eta b}{1+\tilde\Delta c(b)} 
%{1\over 2}\left({\tilde U\over   \pi\tilde\Delta}\right)^2}.
\end{equation}
For $b\to 0$ find $\tilde\Delta c(b)\to {1\over 2}\left({\tilde U\over
    \pi\tilde\Delta}\right)^2=\frac12(R-1)^2$ and so this result corresponds to that of Logan
    and Dickens \cite{LD01}, and as derived from 
the renormalized perturbation  calculations \cite{Hew01}. It shows that the effective
Zeeman splitting of the free quasiparticles, $2\tilde\eta b$, is reduced in the spectral
densities by a factor arising from the $\omega^2$ term in the imaginary part
of the self-energy. In the Kondo limit,  $\tilde U=\pi\tilde\Delta=4T_{\rm
  K}$, so $\tilde U/\pi\tilde\Delta=1$ and $\tilde\eta=R=2$; hence the
peak position of $2b$ is reduced by a factor of $2/3$. 
This effect will be reflected in  the voltage difference between the
differential conductance peaks,
\begin{equation}
e(V_{ds}^{+}-V_{ds}^{-})=\frac{4\tilde\eta b}{1+\tilde\Delta c(b)}
f_c(\tilde\eta b,\tilde\Delta),   
\end{equation}
where $f_c$ is a correction factor due to the overlap of resonances similar
as above, which can be computed numerically. 
However,  in calculating this voltage
difference we cannot neglect the voltage dependence of the self-energy.\par

When we take the $V_{ds}$ dependence of the retarded Green's function into account,
\begin{equation}
{dI\over dV_{ds}}=-{e^2\Delta\over h}
\sum_\sigma{\rm Im}G_{d\sigma}^r(eV_{ds}/2,V_{ds})+{2
e\Delta\over h}\sum_{\sigma}\int_{0}^{eV_{ds}/2}{d\omega}
\left[-{\rm Im}{\partial G_{d,\sigma}(\omega,V_{ds})\over\partial
    V_{ds}}\right].
\label{dc}
\end{equation}
In considering the contribution from the first term on the right hand side we can replace the
$eV_{ds}$ in the self-energy by $2\omega$, and then see where 
it has a maximum as a function of $\omega$. This will  correspond to the
calculation we have just done 
except that the $\omega^2$ term in $\tilde\Sigma_I(\omega)$ will have an extra
factor of $4$, as we can see from equation (\ref{self1}). We get the modified result,
\begin{equation}
e(V_{ds}^{+}-V_{ds}^{-})={4\tilde\eta b\over 1+4\tilde\Delta c(b)}f_c(\tilde\eta b,\tilde\Delta).
\end{equation}
In the Kondo regime and for $b\to 0$ the reduction factor arising from the self-energy
including the voltage dependence is now $1/3$, so the splitting on including  this term
is reduced by an extra factor of $2$ when the voltage dependence in the retarded Green's function is
taken into account.
We do need to take account, however, of the second term on the right hand side
of equation (\ref{dc}) involving an integral over $\omega$. This term can be
written as  
\begin{equation}
{3c(b)e^2\over h}V_{ds}\tilde\Delta \sum_{\nu}\int_0^{eV_{ds}/2}d\omega\,
{(\omega+\nu\tilde\eta b)^2-(\tilde\Delta+c(b)(\omega^2+{3(eV_{ds})^2/ 4}))^2\over
  [ (\omega+\nu\tilde\eta
  b)^2+(\tilde\Delta+c(b)(\omega^2+{3(eV_{ds})^2/4}))^2]^2}.
\label{int}
\end{equation}

\begin{figure}
\begin{center}
\includegraphics[width=0.6\textwidth]{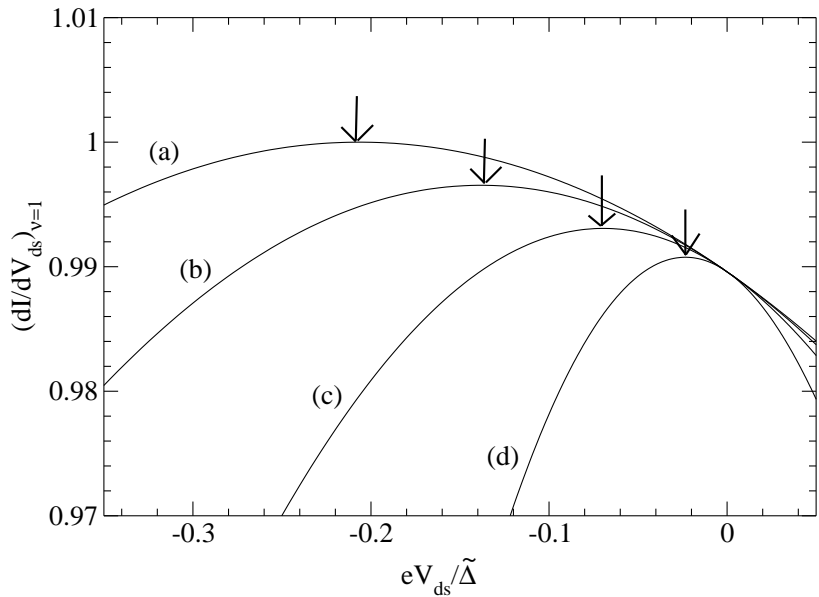}
\smallskip
\end{center}
\caption{The shift of the component resonance ($\nu=1$)  in
  the differential conductance (units of $e^2/h$) in a magnetic field  for
$b/\tilde\Delta =0.05$ as a function of the bias voltage $eV_{ds}/\tilde\Delta$, according
  to the inclusion of different contributions as described in the text. The
  arrows indicate the respective maxima.}
\label{shiftsingspec}
\end{figure}
If we plot all the contributions to $dI/dV_{ds}$ in the very weak field regime
then, due to overlap, no magnetic field splitting can be observed. We can calculate, however,
the shifts in the component resonance for $\nu=\pm 1$. In figure \ref{shiftsingspec} we
plot the terms in the  differential conductance (in units of $e^2/h$) given by
equation (\ref{dc}) as a function of $eV_{ds}/\tilde\Delta$, where  we use
$\nu=1$ in (\ref{ft}) for the first term in (\ref{dc}) and (\ref{int}) for the second
term in (\ref{dc}).  We  take
values corresponding to the Kondo regime, with $R=\tilde\eta=2$, $b/\tilde
\Delta=0.05$ ($\pi\tilde\Delta=4T_{\rm  K}$).
We have distinguished between the different contributions, (a) is the case
for the non-interacting quasiparticles, (b) includes the $\omega^2$ term
 in the renormalized
self-energy, (c) includes the first
term in equation (\ref{dc}), and (d) takes into account the full expression
including the  integral term in   (\ref{int}).
We see that the integral term arising from the voltage dependence of
$G_{d,\sigma}(\omega,V_{ds})$ causes a significant further reduction of the
magnetic shift beyond that estimated from the first term in equation
(\ref{dc}).
In an experimental conductance measurement, this is not observable, however,
due to the overlap of the two components.
\par
%\vspace{3cm}
\begin{figure}
\vspace{1cm}
\begin{center}
\includegraphics[width=0.6\textwidth]{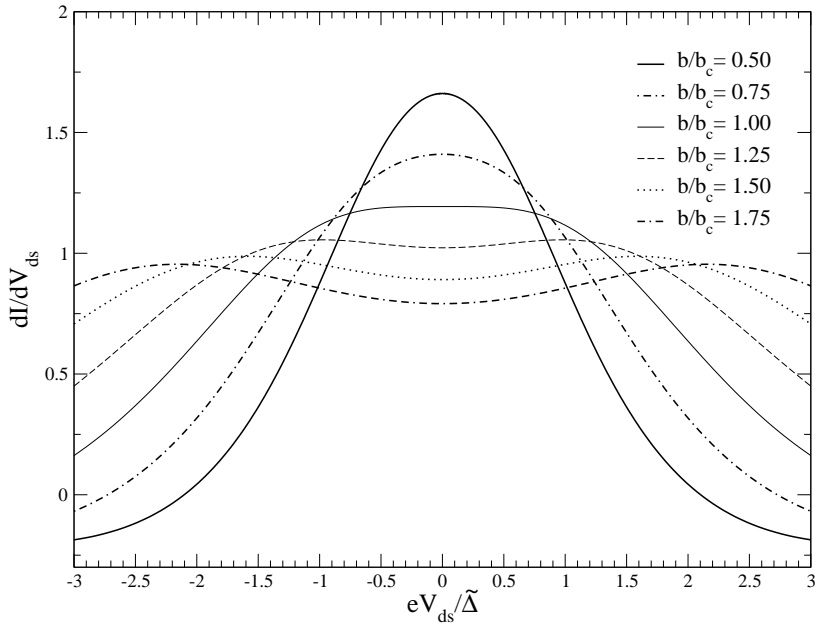}
\smallskip
\end{center}
\caption{The total differential conductance (in units of $e^2/h$) in the Kondo regime
for larger magnetic field values, calculated using equation (\ref{dc}) taking
into account the full selfenergy expansion from (\ref{selfen}). These results
are asymptotically exact for ${eV_{ds}}/{\tilde \Delta}\ll 1$ and
  approximate, based on a second order expansion in $eV_{ds}$ for larger values.}
\label{dIdVtot}
\end{figure}
\section{Critical field}
Our derivation is restricted to the regime where $eV_{ds}$ is small compared
to $\tilde\Delta$. These results are sufficient for us to deduce
the critical value of the magnetic field $b_c$ at which  two distinct peaks
 begin to appear in the total
differential response. For values of $b<b_c$ the differential
conductance will have a maximum at $V_{ds}=0$, and for $b>b_c$ this will
become a minimum. From the coefficient of $dI/dV_{ds}$ to order 
$V_{ds}^2$ we can determine the point at which it changes sign as a function
of $b$, and hence  determine $b_c$.  The contribution from the integral term
in equation (\ref{int}) to order $V_{ds}^2$  can be evaluated trivially, as it is
sufficient to this order
to put $\omega=V_{ds}=0$ in the integrand. 
As a first estimate using the above results the value of  $b_c$
can be calculated analytically, and the result
expressed entirely in terms of $\tilde\Delta$ and the Wilson ratio
$R=\eta(0)=1+\tilde U/\pi\tilde\Delta$, 
\begin{equation}
{b^2_c\over\tilde\Delta^2}={\sqrt{9+20(R-1)^2(1+5(R-1)^2)}-3 \over
  10R^2(R-1)^2}.
\label{critb1}
\end{equation}
In the non-interacting case, $R=1$ and $b_c/\Delta=1/\sqrt{3}=0.577$, which
corresponds to $f_c(\tilde \eta b,\tilde \Delta)=0$ from equation (\ref{fc}),
and in the Kondo regime, $R=2$, $\tilde\Delta=4T_{\rm K}/\pi$, and $b_c/T_{\rm
  K}=0.582$, with $T_{\rm K}$ given by
\begin{equation}
T_{\rm K}=\Delta\left({U\over
  2\Delta}\right)^{1/2}e^{-\pi U/8\Delta+\pi\Delta/2U}.
\end{equation}
If the $V_{ds}$ dependence of the Green's
function is neglected the result in the Kondo regime is  $b_c/T_{\rm K}=0.491$; significantly
smaller than if this term is included.\par
The estimated critical magnetic field is comparable with
$\tilde\Delta$, and for $U\ne 0$ it may not be sufficient to work to linear
order
in $B$. 
It is possible to work with an  arbitrary magnetic field, but in this case
the renormalized parameters become field dependent. The renormalized
self-energy to order $\omega^2$ and $V_{ds}^2$ can be expressed in 
the form,    
\begin{equation}
\tilde\Sigma_\sigma(\omega,V_{ds})=
%-\frac{\pi\tilde U^2(b)\tilde\rho_{d,\sigma}^3(0,b)}{2}
-c(b)  \left[i\left(\omega^2 +3\left(\frac{eV_{ds}}{2}\right)^2\right)+
\frac{\sigma
  b\tilde\eta(b)}{\tilde\Delta(b)}\left(\alpha_\omega(b)\omega^2+\alpha_V(b)\left({eV_{ds}\over   
    2}\right)^2\right)\right],
\label{selfen}
\end{equation}
where $c(b)$ is given in equation (\ref{cb}).
The coefficients for the expansion of the real part of
$\tilde\Sigma_\sigma(\omega,V_{ds})$  are 
\begin{equation}
\alpha_\omega(b)=2+{2I(b)\tilde\Delta(b)\over\tilde\xi(b)[\tilde\rho^{\scriptscriptstyle 0}_{d}(0,b)]^2},
\end{equation} 
where $\tilde\xi(b)=\pi\tilde\rho^{\scriptscriptstyle 0}_{d}(0,b)\tilde\eta(b) b$,
$I(b)$ is the integral 
\begin{equation}
I(b)=\int\limits_{-\infty}^{\infty} 
\int\limits_{-\infty}^{\infty}G^0_{\downarrow}(\omega'')G^0_{\downarrow}(\omega''+\omega')
[G^0_{\uparrow}(\omega')]^3{d\omega''\over 2\pi}{d\omega'\over 2\pi} ,
\end{equation}  
where $[G^0_{\sigma}(\omega)]^{-1}=\omega+\sigma\tilde\eta b
+i\tilde\Delta{\rm sgn}(\omega)$ is the renormalised free propagator for
$T=0$. 
$\alpha_\omega(b)$ is obtained from the second order derivative of the
renormalized self-energy evaluated at $\omega=0$ using the renormalized
perturbation theory to order $\tilde U^2$. This reduces to the exact result of
Yamada \cite{Yam75} for the symmetric model for $b=0$.
The corresponding coefficient $\alpha_V(b)$ for finite voltage is calculated from the $\tilde U^2$
contribution to the retarded self-energy in the Keldysh formalism \cite{Kel65}, where
the propagator $G_{0,\sigma}^{--}(\omega,V_{ds})$ taken to order $V_{ds}^2$
for $T=0$ is
\begin{equation}
  G_{0,\sigma}^{--}(\omega,V_{ds})=G^0_{\sigma}(\omega)-
\frac{i\tilde\Delta\delta'(\omega)(eV_{ds})^2/4}{(\omega-\sigma\tilde\eta  
  b)^2+\tilde\Delta^2},
\end{equation}
where $\delta'(\omega)$ is the derivative of the delta-function, together with
a similar equation for $G_{0,\sigma}^{+-}(\omega,V_{ds})$. The result for
$\tilde\alpha(b)$ is
\begin{equation}
\tilde\alpha_V(b)=1+\frac{\tilde\Delta(b)}{2\tilde\xi(h)\tilde\eta(b)b}
\left[1-\frac{\tilde\eta(b)b}{\tilde\Delta(b)}{\rm 
    tan}^{-1}\left(\frac{\tilde\eta(b)b}{\tilde\Delta(b)}\right)
 \left(4+\frac{\tilde\Delta(b)}{\tilde\xi(h)\tilde\eta(b)b}\right)\right]. 
\end{equation} 
The contribution to the real part of the self-energy to order
$\omega^2$ and $V_{ds}^2$ is proportional to $\sigma \tilde\eta(b)b$. It might be
thought that such a term should cancel out in taking the sum over the two spin components.
However, there is a $\sigma$-independent contribution from a cross term with the effective
Zeeman term $\sigma b\eta(b)$, which has to be included. In the limit $b\to 0$
equation (\ref{selfen}) reduces to (\ref{self1}).
\par
The equation for the critical field $b_c$ becomes
\begin{equation}
  {b^2\over\tilde\Delta^2(b)}={\sqrt{(3-\alpha(b)\gamma(b))^2+4\gamma(b)(5-\alpha(b))(1+5\gamma(b))}-3 
+\alpha(b)\gamma(b)\over 2\gamma(b)(5-\alpha(b))\tilde\eta(b)^2}   
\label{critb2}
\end{equation}
where $\alpha(b)=\alpha_\omega(b)+\alpha_V(b)$, and 
\begin{equation}
\gamma(b)=\pi\tilde\Delta(b)\tilde
U^2(b)[\tilde\rho_d^{\scriptscriptstyle 0}(0,b)]^3=\pi\tilde\Delta(b)\tilde\rho^{\scriptscriptstyle 0}_d(0,b)(R(b)-1)^2.
\end{equation}
%as the Wilson ratio $R$ in a magnetic field is given by $R(b)=1+\tilde U(b)\tilde\rho_d(0,b)$.
Equation (\ref{critb2})  is an implicit equation for $b_c$ which 
can be solved by iteration starting from the much simpler result
(\ref{critb1}), obtained within the linear approximation.\par

For a strong coupling situation ($U/\pi\Delta=4$) the result for the critical
field obtained by iterating equation (\ref{critb2}) 
and using the $b$-dependent renormalized parameters taken from figure \ref{renpar} is
$b_c\simeq 0.459\tilde\Delta=0.584T_{\rm K}$. This differs only by 0.3\% from the value
obtained from (\ref{critb1}). The small difference is due to the
fact that the various correction terms due to the $b$ dependence of the
parameters in the more general formula (\ref{critb2})
tend to cancel giving only a small resultant change. 

Plots of the total differential conductance for various fields above and below
the critical field are displayed in figure \ref{dIdVtot}. We have taken the
full self-energy as given in (\ref{selfen}) into account, 
including the field dependence of the renormalized parameters.
Our results are asymptotically exact only for small $V_{ds}$ and a more
complete theory is required to calculate the splitting at larger bias voltages.
The major problem to be solved is the dependence of the self-energy on the
voltage bias term, when $eV_{ds}$ is comparable with and greater than the
Kondo temperature $T_{\rm K}$, so that a detailed comparison with experiment
can be made with the experimental results in this regime.  
\section{Conclusions}\par
Measurements of the differential conductance in a magnetic field have been
used to infer  the magnitude of the magnetic  splitting of a Kondo resonance
\cite{KAGGKS04,AGKK04}. The results reveal an apparent disagreement with
theoretical predictions, but the comparisons with theory have been based on
calculations of the equilibrium Green's function for the dot. It is not clear
that this will constitute a reasonable approximation at the finite bias voltages
used in the experiment. To examine this question we
 have derived an expression for the differential
conductivity through a quantum dot, described by a symmetric Anderson model,
for small bias voltages and arbitrary magnetic field.
This has enabled us to estimate the effect of the voltage dependent terms 
from the non-equilibrium dot Green's function.  Our estimates of the 
shifts of the component resonances for small magnetic
field values  differ significantly from the values obtained 
using the equilibrium Green's function. We conclude that it
is important to take this voltage dependence into account and use the non-equilibrium
Green's function for a meaningful comparison with experiment.  \par 
Though our calculations are restricted to the small voltage regime 
we have also been able to estimate the value of the critical field $B_c$ for
the emergence of two distinct peaks in the total differential conductance. 
Our approach is applicable in all ranges of the interaction $U$ from the weak
coupling to the strong coupling Kondo regime for $T=0$. They complement the
perturbation calculations in $U$ of Fuji and Ueda 
\cite{FU05} for the same model. It is difficult to make a 
comparison with their results, however, because the interaction simultaneously 
modifies the renormalized parameters, such as $\tilde\eta(b)$,  and introduces
a dependence on $V_{ds}$, both directly and through  the $\omega$ dependence
in the self-energy.  These three effects of the interaction can be clearly
distinguished within our formulation and taken separately into account.
It should be possible to extend our calculations to larger
values of $V_{ds}$  using a renormalized version of the perturbation theory,
and this approach is currently being investigated. 
 \bigskip\par
  
\noindent{\bf Acknowledgement}\par
\bigskip

  ACH wishes to thank the EPSRC for support through the Grant GR/S18571/01).
 JB thanks the Gottlieb Daimler- and Karl Benz-Foundation and EPSRC for financial support, and
AO acknowledges the support by the Grant-in-Aid
for Scientific Research for JSPS. We also wish to thank W. Koller for helpful
discussions. 

\par
%\bibliography{artikel,biblio1}

\end{document}